# Nano-manipulation of diamond-based single photon sources

E. Ampem-Lassen[1], D. A. Simpson[1*], B. C. Gibson[1], S. Trpkovski[1], F. M. Hossain[1], S. T. Huntington[1], K. Ganesan[2], L. C. L. Hollenberg[1], and S. Prawer[1]

[1]*Quantum Communications Victoria, School of Physics, The University of Melbourne, Parkville, 3010, Australia.*
[2]*Micro-Analitical Research centre School of Physics, The University of Melbourne, Parkville, 3010, Australia.*
[*]*Corresponding author:* simd@unimelb.edu.au

**Abstract:** The ability to manipulate nano-particles at the nano-scale is critical for the development of active quantum systems. This paper presents a new technique to manipulate diamond nano-crystals at the nano-scale using a scanning electron microscope, nano-manipulator and custom tapered optical fibre probes. The manipulation of a ~ 300 nm diamond crystal, containing a single nitrogen-vacancy centre, onto the endface of an optical fibre is demonstrated. The emission properties of the single photon source post manipulation are in excellent agreement with those observed on the original substrate.

**1. Introduction**

Diamond has a range of readily produced optical centres that can be utilised as single photon sources (SPSs). Diamond based SPSs offer the unique advantage of long-term photo stability at room temperature which has generated significant interest in areas such as quantum information processing (QIP) [1-3]. Among the single photon emitters identified to date [4-9], the nitrogen vacancy (N-V) centre remains the most studied and applicable for QIP. The N-V centre has been identified in natural diamonds and can also be created synthetically using the microwave plasma enhanced chemical vapour deposition (MPECVD) technique [10] or through direct ion implantation [11]. The ability to readily fabricate these single defect centres has in recent times seen the commercialization of the first SPS based on the N-V defect in diamond [12]. However, fabricating these sources in desired locations and on particular structures remains difficult due to the statistical nature of creating the atomic defect. The demonstration of nano-scale magnetometry [13-16] and the possibility of de-coherence imaging using single N-V centres opens up exciting opportunities in quantum imaging (QI) [17]. However, if these types of opportunities are to be explored the manipulation of single quantum systems and sources is required. A previously reported method has been used to manipulate nano-particles containing single photon emitters by combining an atomic force microscope (AFM) with a scanning confocal optical microscope [13]. This technique allows the position and optical characteristics of the emitter to be measured simultaneously provided the focused laser spot from the confocal microscope coincides with the AFM cantilever tip. Combined AFM and confocal systems have been shown to be an effective tool in terms of attaching single quantum emitters onto the AFM tip. However, the technique is limited to substrates which are transparent and optically thin.

In this paper we present a technique for characterizing and manipulating isolated diamond nano-particles containing single photon emitters. The nano-manipulation of single diamond nano-crystals is achieved using a custom tapered optical fibre probe in a standard scanning electron microscope (SEM) with a nano-manipulator. The technique offers the unique advantage of being able to inspect and manipulate single diamond nano-crystals onto desired substrates and waveguiding devices with nano-scale precision.

## 2. Experimental details

To correlate the position of a single diamond nano-crystal containing a single defect centre in the SEM environment a marked silicon substrate was prepared using a focused ion beam (FIB). The markers were milled using 30 keV $Ga^+$ ions with a current of 7 nA to a depth of ~ 3 µm to enable detection in the confocal microscope. The marked silicon substrate was then ultrasonically seeded in a solution of methanol and commercial grade diamond powder (0-500 nm) known to contain appreciably high numbers of (N-V) defect centres. The diamond nano-particles and subsequently rinsed in acetone, methanol and de-ionised water to remove the unwanted larger and coagulated particles.

The marked region of the silicon substrate was characterised optically with an in-house scanning confocal microscope. The room temperature luminescent properties of the emitting centres were studied under continuous wave (CW) excitation at 532nm with a 100× 0.95 NA microscope objective, providing a spatial resolution of ~ 400nm. The photo-luminescence from the emitting crystals captured by the microscope objective was coupled into a standard 62.5/125 µm multimode optical fibre which acted as a pinhole to provide the confocallity of the microscope. A 560nm long-pass and 650-750nm band-pass filter was used to remove the unwanted pump excitation before being coupled into the fibre. The photon statistics of the diamond emitters were characterised using a fibre based version of the Hanbury Brown and Twiss (HBT) interferometer [18]. The characteristic dip in the second order correlation function at zero delay time enabled single N-V centres to be identified. The confocal images of the marked silicon substrate were then used to correlate the position of the single N-V centres in the SEM. The SEM was operated with a nano-manipulator containing a custom tapered optical fibre probe with a tip size of ~ 50nm to manipulate the diamond nano crystals [19]. The probe was fabricated using a custom $CO_2$ laser based fibre pulling system.

## 3. Results and discussion

The FIB marked silicon substrate used to correlate the position of the single photon emitters is shown in Fig. 1.

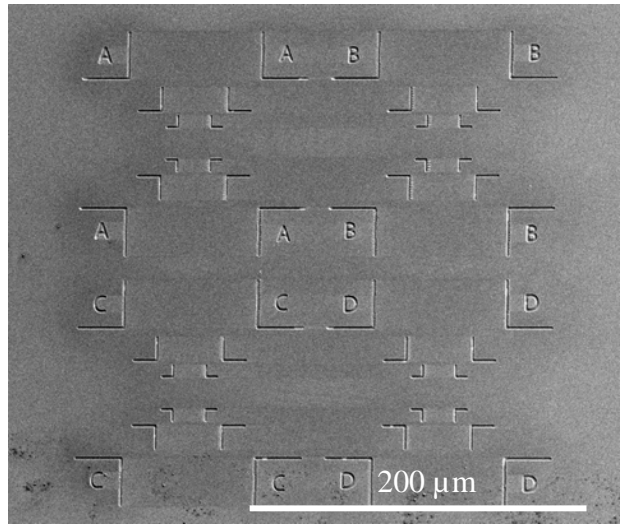

Fig. 1: Focused ion beam (FIB) marked silicon substrate prior to diamond seeding. Note: the depth of the milled markers was ~ 3 µm.

A fluorescence intensity confocal map of a 70×70 µm region of the seeded marked substrate is shown in Fig. 2(a). The photon statistics of each fluorescent emitter were studied to identify a single N-V centre. The fluorescent emitter highlighted in Fig. 2(a) was found to exhibit the characteristic dip in the second order correlation function at the zero delay time (t=0) (see Fig. 2(d)), with a minimum $g^{(2)}(0)$ of 0,16 observed under 500 µW of excitation power at 532 nm. The measured single photon emission rate was ~ 270 kHz under the same excitation conditions. After the initial optical characterization, the sample was placed into the SEM for inspection and manipulation. The properties of silicon are such that the substrate does not experience significant charging in the SEM when operated with 10 keV electrons at 50 pA. The fabricated optical probe however does suffer from charging due to insulating nature of silica. Therefore, to avoid the charge related issues, the optical probe was coated with a thin layer of carbon ~ 10 nm. The areas of the marked substrate were compared to the confocal images to correlate the position of the single N-V emitter. Figure 2(b) shows the SEM image of a marked region of the substrate.

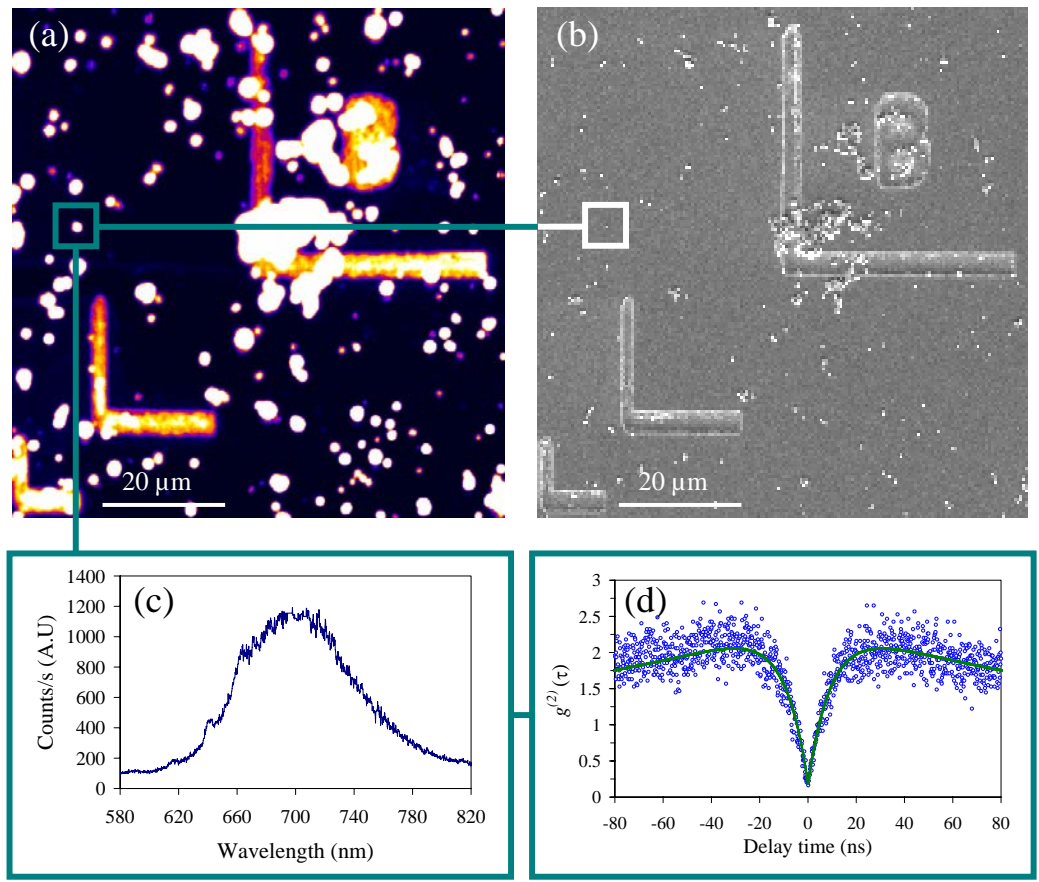

Fig. 2: (a) Confocal map of a 70×70 µm area of the marked and seeded silicon substrate. (b) SEM image of the same 70×70 µm region of a marked and seeded silicon substrate (c) The photo-luminescence spectrum of the fluorescence emitter identified in (a). (d) The photon statistics of a single N-V centre identified in (a) measured at room temperature with 500 µW excitation power at 532 nm.

Once the position of the single emitting centre was identified, the tapered probe was manipulated to pick the diamond nano-crystal from the substrate. A strong electrostatic attraction between the tip and diamond nano-particle firmly holds the particle in place allowing the probe to be re-positioned to another location where the particle is required to be placed. It is worth mentioning here that several steps had to be taken to manipulate the single emitter onto a waveguiding structure in the form of a single mode optical fibre (3M FS-SN 3224). Firstly, in order to identify the core region of an optical fibre in the SEM, the end face was etched in 25% hydrofluoric (HF) acid for 2 minutes [20]. Secondly, the bulk of the optical fibre is made up of $SiO_2$ which suffers from significant charging in the SEM, to eliminate these effects the optical fibre was coated with a thin layer of carbon ~ 10 nm. The emitter was manipulated to a region outside the core to avoid the unwanted fluorescence background from co-dopants such as germanium and fluorine [20]. Figures 3(a)-(c) illustrate a sequence of SEM images taken before and after the manipulation of the SPS, with the corresponding confocal image of the endface of the optical fibre shown in Fig. 3(d).

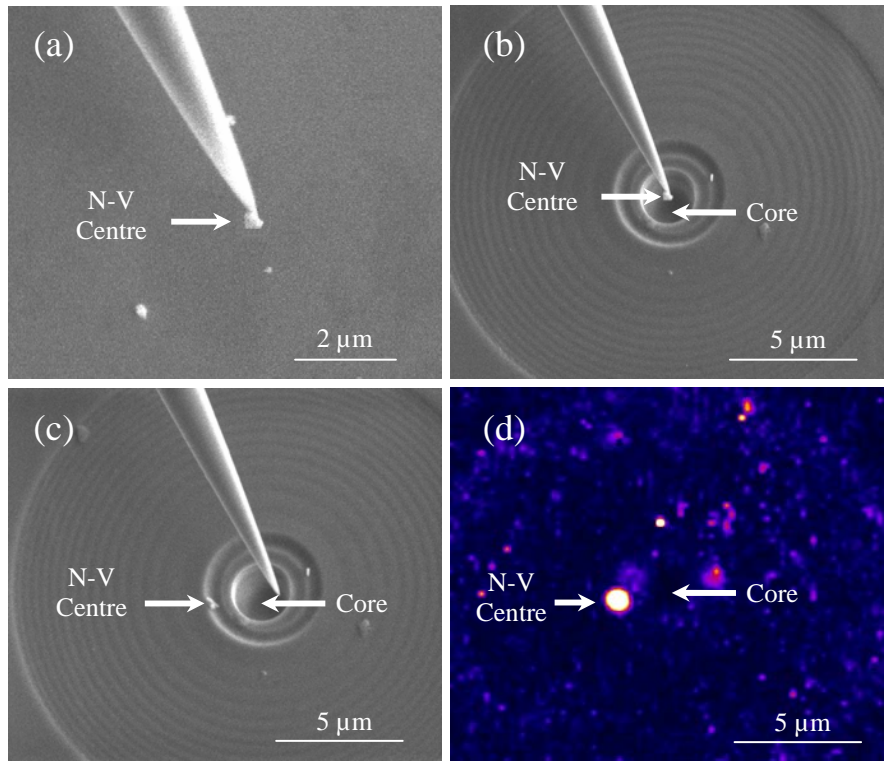

Fig. 3: (a) SEM image of a single diamond nano-crystal (~ 300 nm) being removed from the marked silicon substrate. (b) SEM image of the tapered probe with the diamond nano-crystal attached and positioned above the optical fibre endface. (c) SEM image of the single emitting diamond nano-crystal positioned in the region outside the fibre core. (d) Confocal image of the fibre endface after manipulation and 1 hour anneal at 650 °C in air.

After manipulation, the thin layer of carbon coating applied to the endface of the optical fibre the sample was removed by annealing the fibre at 650 °C in air using an 1100 Quartz Tube furnace for 1 hour. To confirm that the manipulated N-V centre was not damaged or degraded in the process, the centre was characterized again using the same experimental configuration and compared with the results in Fig. 2 obtained on the silicon substrate. A confocal map of the post annealed fibre endface is shown in Fig. 3(d) which clearly identifies the single N-V emitter outside the core region of the fibre. It is worth mentioning here that the fluorescence from the core region of the fibre is not observed in the confocal map as the etched region of the core is beyond the depth of field (DOF) of the microscope objective. Although perhaps the most striking feature of the confocal image is the contrast between the SPS and its surroundings. In the 17 x 17 µm scan, the single N-V centre is the only notable fluorescing object, allowing it to be easily identified. The photon statistics from the single centre after manipulation are shown in Fig. 4(a) and the measured single photon emission rate as a function of incident pump power is shown in Fig. 4(b). The performance characteristics of the centre are in excellent agreement with those obtained on the original silicon substrate. The autocorrelation function minimum at the zero delay time, $g^{(2)}(0)$, was measured to be 0.16 before and after manipulation and the measured single photon emission rate was within 5% of the rate obtained on the original substrate. These results confirm that the manipulation technique does not degrade the performance of the single photon source.

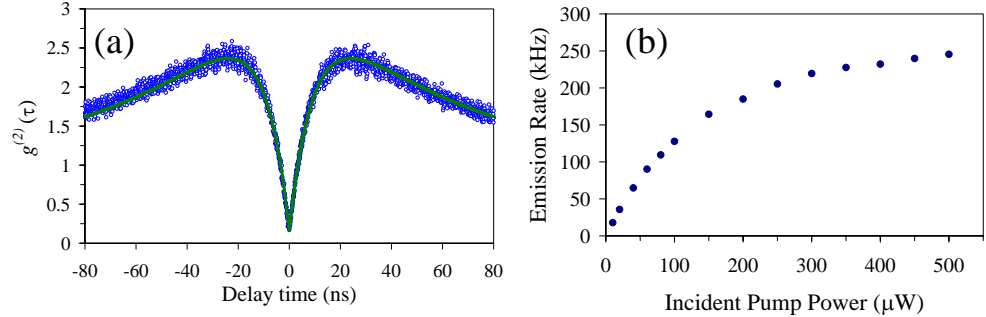

Fig. 4: (a) The photon statistics of the single N-V centre on the endface of the optical fibre measured at room temperature with 500 µW excitation power at 532 nm. (b) Single photon emission rate as a function of incident pump power.

This technique opens up the possibility to manipulate single photon sources onto structures such as mirrors, nano-wires and cavities which can all act to enhance the emission properties [21-23]. Furthermore, the ability to nano-manipulate single quantum systems could prove advantageous in realizing QIP and QI applications. Although the sample preparation can be laborious, economies of scale can be achieved since a single seeded substrate contains numerous single defect N-V centres.

**4. Conclusion**

A novel technique for manipulating single diamond nano-crystals, containing single N-V defect centres, was presented. The technique maintains the performance characteristics of a single N-V centre in a 300 nm diamond crystal before and after manipulation. The ability to inspect the nano-particles throughout manipulation process provides an important adjunct to existing methods for analyzing and manipulating single photon sources. Manipulating these types of quantum systems onto nanostructures and other plasmonic and cavity devices is the goal of future work.

**Acknowledgements**

The authors acknowledge Dr. Sergey Rubanov for helpful discussions regarding the operation of the FIB/SEM. This project was supported by Quantum Communications Victoria, which is funded by the Victorian Government's Science, Technology and Innovation initiative.